\documentclass[aps,prl,twocolumn,showpacs,preprintnumbers,amsmath,amssymb]{revtex4-1}
\usepackage{graphicx}
\usepackage[usenames,dvipsnames]{xcolor}
\usepackage{braket}
\usepackage{float}
\usepackage{lineno}
\usepackage{multirow}
\begin{document}
\title{Quantum simulation of the universal features of the Polyakov loop}
\author{Jin Zhang$^{1}$}
\author{J. Unmuth-Yockey$^2$}
\author{J. Zeiher$^3$}
\author{A. Bazavov$^4$}
\author{S.-W. Tsai$^1$}
\author{Y. Meurice$^5$}
\affiliation{$^1$Department of Physics and Astronomy, University of California, Riverside, CA 92521, USA}
\affiliation{$^2$Department of Physics, Syracuse University, Syracuse, NY 13244, USA}
\affiliation{$^3$ Max-Planck-Institut f\"ur Quantenoptik, 85748 Garching, Germany}
\affiliation{$^4$Department of Computational Mathematics, Science and Engineering,
and Department of Physics and Astronomy,
Michigan State University, East Lansing, MI 48824, USA}
\affiliation{$^5$Department of Physics and Astronomy, The University of Iowa, Iowa City, IA 52242, USA}
\definecolor{burnt}{cmyk}{0.2,0.8,1,0}
\def\lt{\lambda ^t}
\def\note{note}
\def\beq{\begin{equation}}
\def\enq{\end{equation}}
\newcommand{\Tr}{\text{Tr}}

\date{\today}
\begin{abstract}
Lattice gauge theories are fundamental to our understanding of high-energy physics. Nevertheless, the search for suitable platforms for their quantum simulation has proven difficult. 
We show that the Abelian Higgs model in 1+1 dimensions is a prime candidate for an experimental quantum simulation of a lattice gauge theory.
To this end, we use a discrete tensor reformulation to smoothly connect the space-time isotropic version used in most numerical lattice simulations to the continuous-time limit corresponding to  the Hamiltonian formulation. 
The eigenstates of the Hamiltonian are neutral for periodic boundary conditions, but we probe the nonzero charge sectors by either introducing a Polyakov loop or an external electric field. 
In both cases we obtain universal functions relating the mass gap, the gauge coupling, and the spatial size which are invariant under the deformation of the temporal lattice spacing. 
We propose to use a physical  multi-leg ladder of atoms 
trapped in optical lattices and interacting with Rydberg-dressed interactions
to quantum simulate the model and check the universal features. 
Our results provide a path to the analog quantum simulation of lattice gauge theories with atoms in optical lattices.
\end{abstract}

\maketitle
Lattice gauge theories (LGT) are fundamental to our understanding of strongly interacting particles in high energy physics. 
Translating the success of quantum simulations with cold atoms in optical lattices~\cite{Gross2017} of systems relevant to condensed matter such as the Bose-Hubbard model to the quantum simulation of LGT would open the door to real-time and finite-density calculations which are beyond the realm of classical computations. An important first step is to achieve this goal for models in one space and one time (1+1) dimensions. 
While the dynamics of the Schwinger model, quantum electrodynamics in 1+1 dimensions, has been explored using a few qubit digital quantum simulation in a system of trapped ions \cite{martinez2016real} or classical-quantum algorithms on 
IBM quantum computers \cite{Klco:2018kyo}, the analog quantum simulation of gauge theories with cold atoms requires complex experimental settings. Existing efforts involve mixtures of bosonic and fermionic atoms \cite{PhysRevLett.109.175302,Kasper:2016mzj} or dipolar interactions of cold molecules \cite{gadway} and are still in progress. 

In this Letter, we propose to quantum simulate the Abelian Higgs model in 1+1 dimensions, the Schwinger model with the electron replaced by a complex scalar field.
The validity of the proposal can be checked by measuring the Polyakov loop, an observable that plays an important role in finite-temperature studies \cite{ploopreview} and for which we present remarkable finite size scaling (FSS) properties. 
We invoke 
a single atomic species in an optical lattice on a multi-leg ladder and recently explored Rydberg-dressed interactions in this platform~\cite{Zeiher2016}. The ladder is a physical lattice with a long side (``legs") representing one-dimensional space and a short side (``rungs"), 
with a shorter lattice spacing, representing rotor angular momentum. 
We aim at maximal simplicity both on the theoretical and experimental side. 
In contrast to other approaches \cite{Zohar:2011cw,Zohar:2012ay,Tagliacozzo:2012vg,PhysRevLett.110.125304,Kasamatsu2013,wiesereview,Zohar:2015hwa,kuno2015real,kuno2017quantum,Cuadra2017,PhysRevLett.109.175302,Kasper:2016mzj,gadway}, we use a manifestly gauge-invariant formulation \cite{PhysRevD.92.076003} where there is no need to enforce Gauss's law. In addition, we consider the limit \cite{PhysRevD.92.076003} 
where the scalar self-coupling becomes large and the Higgs mode  decouples from the low energy theory. We are then left with 
a gauged $O(2)$ spin model with {\it compact} field integration. Fourier analysis provides a {\it discrete} reformulation, suitable for 
quantum simulations, in agreement with Pontryagin duality \cite{pdual}.
\begin{figure}[t]
 \centering
\includegraphics[width=8.6cm]{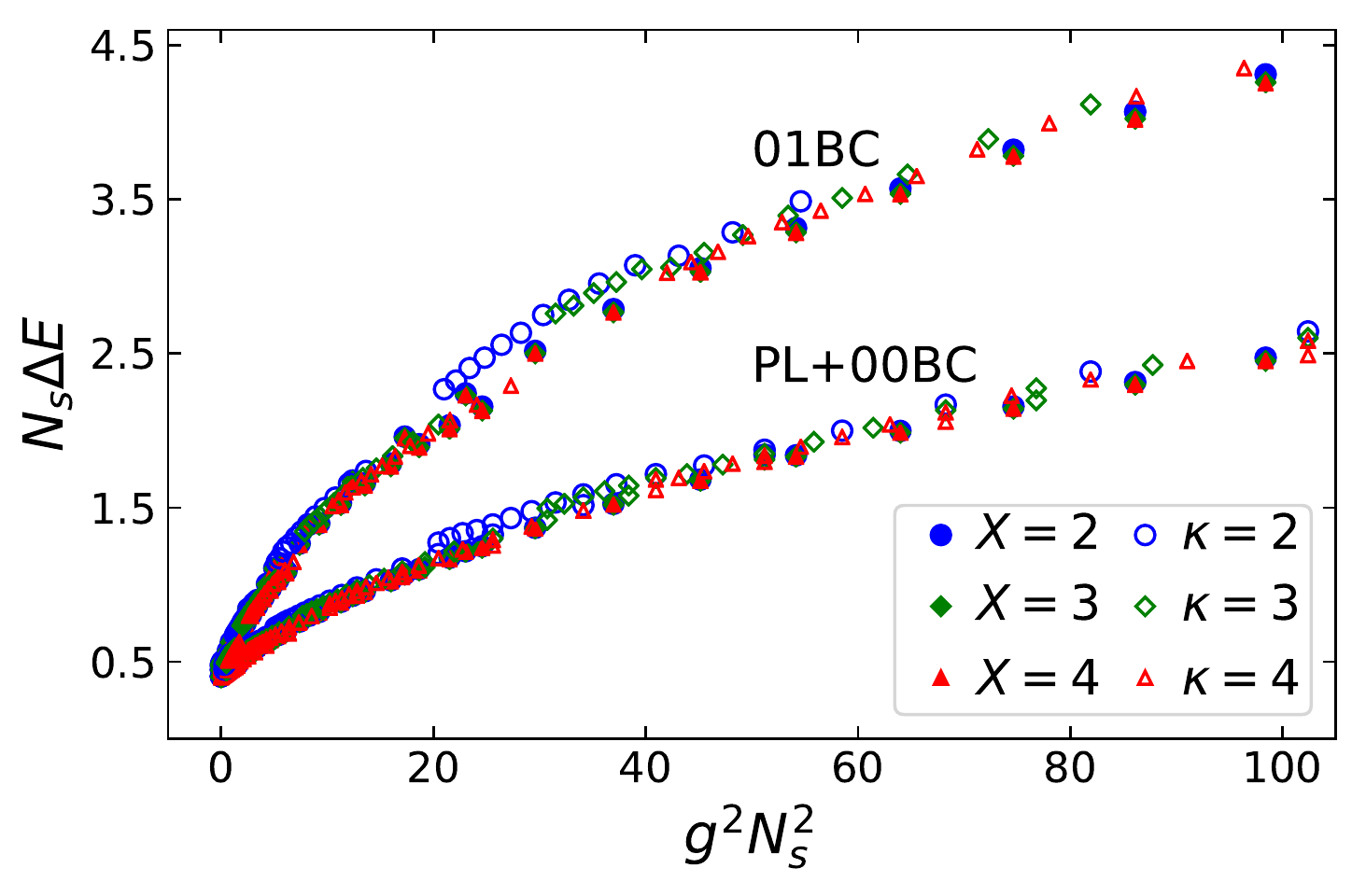}
\caption{\label{fig:collapse1} $N_s\Delta E$ versus $g^2N_s^2$ for the gap $\Delta E$ created by the insertion of the Polaykov loop (lower set) or an external electric field (01 boundary conditions, upper set). Open (filled) markers represent Lagrangian (Hamiltonian) data. 
The choices of parameters, units and methods for both of the 24 datasets are explained in the text.} 
\end{figure} 

The remarkable FSS properties are illustrated by the collapse of 24 datasets in Fig. \ref{fig:collapse1}.
There are four different spatial sizes $N_s$ represented in the figure (4, 8, 16, and 32) and it is 
possible to probe the critical behavior with systems of modest spatial sizes. 
We first give some brief explanations about these results and provide the details later in the text. 
The FSS is related to the energy gap $\Delta E$ created by inserting a Polyakov loop (a Wilson loop wrapping around the periodic Euclidean time) or 
by applying an external electric field. When the gauge coupling $g$ approaches zero, we have an $O(2)$ model and when the hopping parameters exceed their critical value at the  Berezinsky-Kosterlitz-Thouless (BKT) transition, we have infinite correlation length at infinite volume and $\Delta E\propto 1/N_s$ at finite $N_s$. 
Fig. \ref{fig:collapse1} indicates that when we turn on $g$, $N_s \Delta E$ is a linear function of $(g N_s)^2$ at small argument and then a linear 
function of $g N_s$ at larger argument. 
The two parts of Fig. \ref{fig:collapse1} each contain 24 data sets with 12 from a discrete Euclidean time Lagrangian, while the other 12 are from the continuous-time Hamiltonian limit.
It is remarkable that the two calculations provide the same universal functions. Ultimately, it is this equivalence between the two formulations which enables the transfer of results from the experimentally accessible Hamiltonian dynamics to the Lagrangian formulation.

The considered Abelian Higgs model is described by 
the lattice path integral 
$Z=\int D\phi^\dagger D\phi DU e^{-S}$
with action
\begin{eqnarray}
\label{eq:action}
S &=& -\beta_{pl}\sum_x\sum_{\nu<\mu}{\rm Re}{\rm Tr}\left[U_{x,\mu\nu}\right]
\nonumber
\\
&-&\kappa\sum_x\sum_{\nu=1}^{d}
\left[\phi_x^\dagger U_{x,\nu}\phi_{x+\hat\nu}+
\phi_{x+\hat\nu}^\dagger U^\dagger_{x,\nu}\phi_x
\right]. 
\end{eqnarray}
The  
complex (charged) scalar field is $\phi_x={\rm e}^{i\theta_x}$ on space-time sites $x$ 
and the Abelian
gauge fields $U_{x,\mu}={\rm e}^{iA_\mu(x)}$ on the links  from $x$ to  $x+\hat{\mu}$.  
The electromagnetic tensor $F_{\mu\nu} = \partial_\mu A_\nu - \partial_\nu A_\mu$ 
appears when taking  products of gauge fields around an elementary square (plaquette) in the $\mu\nu$ plane.
 The notation for this product is 
$U_{x,\mu\nu}={\rm e}^{i(A_\mu(x)+A_\nu(x+\hat{\mu})-A_\mu(x+\hat{\nu})-A_\nu(x))}$ and the gauge coupling enters through 
$\beta_{pl}=1/g^2$.  The parameter $\kappa$ is the hopping coefficient. 

The Fourier expansions of the Boltzmann weights lead to  expressions of the partition function in terms of discrete sums of products of modified Bessel functions with integer orders on each plaquette and each link of the square lattice. The integer plaquette quantum numbers completely determine  the integer link quantum numbers which are the difference of the neighboring plaquette quantum numbers which  
can be interpreted as {\it dual} variables. Explicit formulas and sign conventions are given in Ref. \cite{PhysRevD.92.076003}, where we also show that the discrete tensor renormalization group (TRG) \cite{PhysRevB.86.045139,prd88} approach and the standard Monte Carlo approach give consistent numerical answers. 
The link quantum numbers can be interpreted as matter charges and 
their sum on the time links between two successive time slices stay constant as time is increased and define a conserved charge $Q$. 
When periodic boundary conditions (PBC), or open boundary conditions with zero taken on the boundaries (00BC), are imposed for the gauge degrees of freedom, the condition $Q=0$ is automatically enforced.

\begin{figure}[t]
 \includegraphics[width=8.6cm]{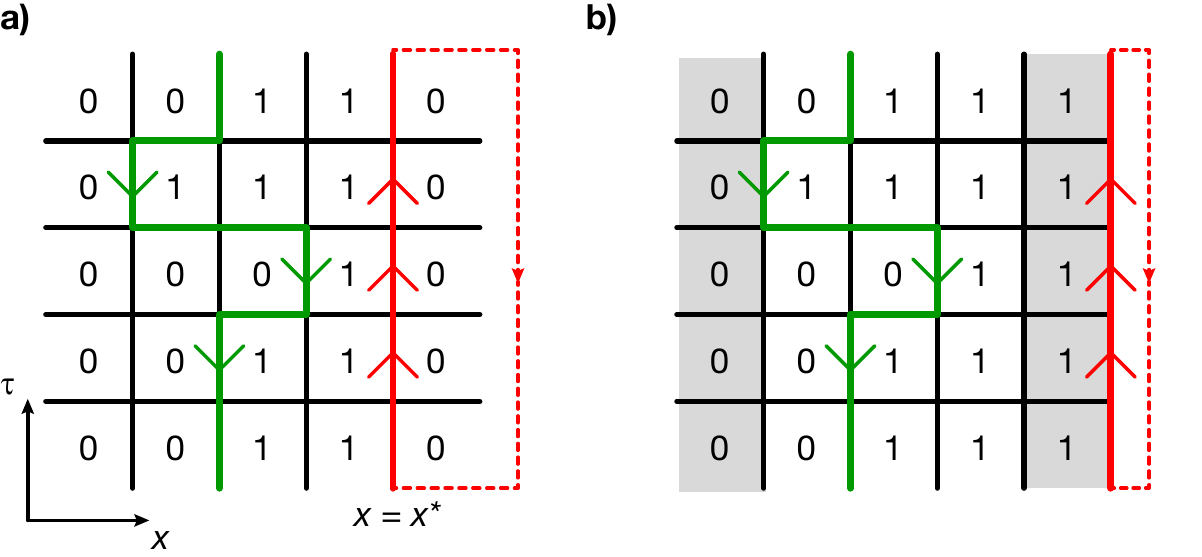}
 \caption{ \label{fig:pandm}The Polyakov loop (arrows pointing up), the matter loop (arrows pointing down) composed of matter charges and plaquette quantum numbers for a) 00BC and b) a situation equivalent to 01BC obtained by ``sliding" the Polyakov loop to the boundary. The dotted lines indicate the wrapping in Euclidean time direction.}
\end{figure}

The $Q\neq0$ sectors can be probed by inserting a product $P$, called the Polyakov loop, of gauge links in the Euclidean time direction $\hat{\tau}$ with periodic boundary conditions: 
\begin{equation}	P = \prod_{n=0}^{N_{\tau}-1} U_{x^{*}+n\hat{\tau}, \hat{\tau}}, 
\end{equation}
inside the path integral. This is illustrated in Fig. \ref{fig:pandm}. Conventional Monte Carlo simulations and TRG 
calculations with a typical bond dimension of $D_{\text{bond}} \approx 40$ provide consistent evidence \cite{prdprogress} for the behavior 
\begin{equation}
\langle P \rangle \propto e^{-N_{\tau} \Delta E}. 
\end{equation}
$\langle P \rangle $ can be interpreted in terms of the free energy induced by the inclusion of a static charge. 
A similar effect can be induced by using asymmetric boundary conditions such as 0 on one side and 1 on the other side 
for the plaquette quantum numbers. 
This situation can be interpreted as the introduction of an external electric field and is denoted as 01BC in Fig. \ref{fig:collapse1} and hereafter.
A similar situation can be accomplished by using 00BC and moving the Polyakov loop to the boundary as shown in Fig. \ref{fig:pandm}b where an addtional column of zeros is implicit.

In order to connect the LGT calculations in the Lagrangian formulation 
with quantum simulations using the Hamiltonian formulation, we need to 
take the time continuum limit. This is done \cite{PhysRevD.92.076003} by taking $\kappa_{\tau}, \beta_{pl} \rightarrow \infty$ while simultaneously taking $\kappa_{s}, a \rightarrow 0$ ($a$ is the temporal lattice spacing) such that the combinations
\begin{equation}
\label{eq:abhamil}
	U \equiv \frac{1}{\beta_{pl} a} = \frac{g^{2}}{a}, \quad
    Y \equiv \frac{1}{2 \kappa_{\tau} a}, \quad
    X \equiv \frac{2 \kappa_{s}}{a}
\end{equation}
are kept constant.  Note that $X$ here is related to $\tilde{X}$ in Ref. \cite{PhysRevD.92.076003} by $X = \sqrt{2} \tilde{X}$.  In this limit, using the properties of the Bessel functions, the Hamiltonian can be identified as
\begin{equation}
	\label{eq:ham}
	H = \frac{U}{2}\sum_{i=1}^{N_s}  \left(L^z_{i}\right)^2 + \frac{Y}{2} {\sum_i}  ' (L^z_{i} - L^z_{i+1})^2-
	X \sum_{i=1}^{N_s} U^x_{i} \ .
\end{equation}
The sum $\sum_i '$, taking 00BC into account, includes $(L^z_{1})^2 + (L^z_{N_s})^2$.
The operator $U^x = (U^+ + U^-)/2$ with the special type of ladder operators $U^+, U^-$ defined by $U^{\pm} \ket{m} = \ket{m\pm1}$,
where $L^z \ket{m} = m \ket{m}$ with plaquette quantum numbers $m = 0, \pm 1, \pm 2, \ldots$ If we truncate at $|m|_{max} = s$, we call it a spin-$s$ truncation even though it is different from the rotation group representation used in Ref. \cite{pra90}.

In the continuous-time limit, the introduction of the Polyakov loop amounts to changing the Hamiltonian into
\begin{align}
\label{eq:plooph}
	\tilde{H} &= \frac{U}{2} \sum_{i = 1}^{N_s} (L_i^z)^2 + \frac{Y}{2} {\sum_{i \neq \frac{N_s}{2}}} ' (L_i^z - L_{i+1}^z)^2 \nonumber \\
    &+ \frac{Y}{2} (L_{\frac{N_s}{2}}^z - L_{\frac{N_s}{2}+1}^z - 1)^2 -X \sum_{i = 1}^{N_s} U_i^x \ ,
\end{align}
where we have assumed that the Polyakov loop is put on the center of the lattice. 
The 01BC choice provides another way to probe the charge-1 sector. 
This simply changes $(L^z_{N_s})^2$ in the second summation of Eq. (\ref{eq:ham}) to $(L^z_{N_s} - 1)^2$.

The numerical continuous-time results were obtained using the density matrix renormalization group (DMRG) \cite{PhysRevLett.69.2863,schollwock2011density} to calculate the ground state energies for both cases. The finite DMRG algorithm with matrix product state (MPS) \cite{PhysRevLett.75.3537} optimization was performed using the ITensor C++ library \footnote{Version 2.1.1, http://itensor.org/}.  $Y = 1$ units were used in all DMRG calculations.

We are now in position to provide more details about Fig. \ref{fig:collapse1}. As explained in more detail in \cite{prdprogress}, arguments regarding the behavior at small and large $gN_s$ led us to conjecture that $N_{s} \Delta E$ is solely a function of the product $(gN_{s})^{2}$.
Fig. \ref{fig:collapse1} supports this idea and shows a good data collapse across multiple system sizes for both the discrete and continuous-time limits. Note that for the discrete-time (Lagrangian) calculations at various $\kappa$, $\Delta E$ was rescaled by 2$\kappa$ while for the continuous time (Hamiltonian)  calculations, no such rescaling was introduced for the different values of $X$.
We emphasize that the collapse is by no means automatic. It breaks down for $\kappa$ not large enough, if we increase $g$ to large values  while keeping $N_s$ constant (there are hints of this in Fig. \ref{fig:collapse1} for the small $N_s$ data), or if the truncation value $m_{max}$ is too small.

Notice that in all cases, 
$N_s \Delta E \simeq 0.5$ when $g^2 \simeq 0$, which corresponds to the gapless BKT phase in the $O(2)$ limit.
We also notice that the energy gap at finite $g^2$ for 01BC is bigger than the gap for PL-00BC. Because 01BC breaks the inversion symmetry of the system, creating a charge on the left costs more energy than that on the right. We can understand this by doing the transformation $L_{i}^{\prime z} = L_{i}^z - 1$ for $i > N_s / 2$. Then the Hamiltonian with 01BC is related to the Hamiltonian with the Polyakov loop by $H_{01} \rightarrow \tilde{H} + U \sum_{i = N_s/2 + 1}^{N_s} L^{\prime z}_i + N_sU / 4$, which is simply adding a linear potential on the right half of the system. But the data collapse is still present, indicating the same universality class. As discussed in Ref. \cite{prdprogress}, a spin-6 truncation is good enough to probe the finite size scaling of the energy gap in the O(2) limit for $2 \leq X \leq 5$ up to $N_s = 32$. As a finite gauge coupling $g$ will suppress the contribution of high plaquette quantum numbers, a spin-6 truncation should work well for any finite $g$ and is used for all DMRG calculations in this work.

\begin{figure}[t]
\includegraphics[width=8.6cm]{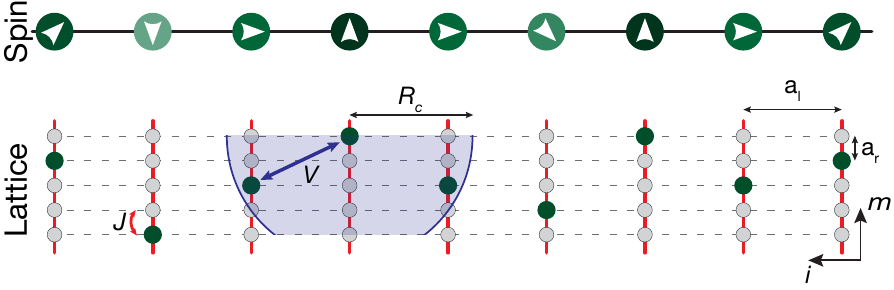}
\caption{\label{fig:ladderpic} Multi-leg ladder implementation for spin-$2$. The upper part shows the possible $m_z$-projections. Below, we show the corresponding realization in a ladder within an optical lattice. The atoms (green disks) are allowed to hop within a rung with a strength $J$, while no hopping is allowed along the legs. The lattice constants along rungs and legs are $a_r$ and $a_l$ respectively. Coupling between atoms in different rungs is implemented via an isotropic Rydberg-dressed interaction $V$ with a cutoff distance $R_c$ (marked by blue shading).}
\end{figure}

An important feature of the Hamiltonians considered above is that $L_i^z$ has positive and negative eigenvalues and cannot 
be realized as the number operator of a Bose-Hubbard model unless a large chemical potential is introduced \cite{pra90,pra96,prd96}. 
For this reason, two species of atoms were introduced in Ref. \cite{pra90}. For a similar reason, a 2-leg ladder with $2s$ atoms per rung  for 
a spin-$s$ truncation was suggested in Ref.  \cite{PhysRevD.92.076003}, however the hopping along a rung can only emulate the $L^x$ operator in the rotation group representation instead of the $U^x$ operator in Eq.~(\ref{eq:ham}) and (\ref{eq:plooph}). Here, we propose a simpler experimental realization to overcome this difficulty, namely 
an asymmetric ladder of $N_s$ rungs of length $2s+1$ each, with lattice constant $a_l$ and $a_r$ along legs and rungs respectively, see Fig.~\ref{fig:ladderpic}. The tunnel coupling along the legs is vanishing while it has a strength $J$ along the rungs. The number of atoms per rung is held fixed at unity, such that the $L^z$-projection of the spin is encoded in the position $m$ of the atom within a given rung and can be read out with near-unity fidelity in a quantum gas microscope~\cite{Gross2017}. The initialization of the system can be achieved in such a setup by preparing an atomic Mott insulator and employing site-resolved optical potentials~\cite{Weitenberg2011a}. 

The Hamiltonian of such a ladder system can be written as 
\begin{align}
\hat{H} = &-\frac{J}{2}\sum_{i=1}^{N_s}\sum_{m=-s}^{s-1}\left( \hat{a}_{m,i}^\dagger \hat{a}_{m+1,i}+h.c\right) - \sum_{i=1}^{N_s}\sum_{m=-s}^{s}\epsilon_{m,i}\hat{n}_{m,i}\nonumber\\
    &+ \sum_{i,i^\prime=1}^{N_s}\sum_{m,m^\prime=-s}^{s}V_{m,m^\prime,i,i^\prime}\,\hat{n}_{m,i}\hat{n}_{m^\prime,i^\prime}.\label{eq:ladderH}
\end{align}
Here, we have additionally introduced an interaction $V_{m,m^\prime,i,i^\prime}$ between two particles at positions $(m,i)$ and $(m^\prime,i^\prime)$ as well as an on-site potential $\epsilon_{m,i}$.
The term $X\sum_iU^x_i$ in Eq.~(\ref{eq:ham}) and (\ref{eq:plooph}) directly maps to the tunneling term in Hamiltonian~\eqref{eq:ladderH} for $J=X$.
Realizing the other two terms requires fine-tuned values $V_{m,m^\prime,i,i^\prime}=V_{m,m^\prime}\,\delta_{i^\prime,i+1}$ with $V_{m,m^\prime}=-|V_0| + Y(m-m^\prime)^2/2$ for the interaction potential between two particles constrained by the Kronecker-symbol $\delta_{i^\prime,i+1}$ to be located in two neighboring rungs. Furthermore, the on-site potentials have to be tuned to $\epsilon_{m,i}=-Um^2/2-(N_s-1)|V_0|/N_s$ for the rungs with $i\neq1,N_s$ and $\epsilon_{m,1}=\epsilon_{m,N_s}=-(U+Y)m^2/2-(N_s-1)|V_0|/N_s$ for two rungs at the boundaries.

Introducing a Polyakov loop amounts to changing the on-site potentials on the two central rungs $N_s/2$ and $N_s/2+1$ to $\epsilon_{m,N_s/2} = -Um^2/2 + Y m - (N_s-1)|V_0|/N_s$ and $\epsilon_{m,N_s/2+1} = -U m^2/2 - Y (m + 1/2) - (N_s-1) |V_0|/N_s$, respectively. The boundary condition 01BC can be realized by tuning the on-site potential at one end of the ladder to $\epsilon_{m,N_s} = -U m^2/2 + Y (m - 1)^2/2 - (N_s-1)|V_0|/N_s$.

\begin{figure}[t]
\vskip-0.3cm
\includegraphics[width=8.6cm]{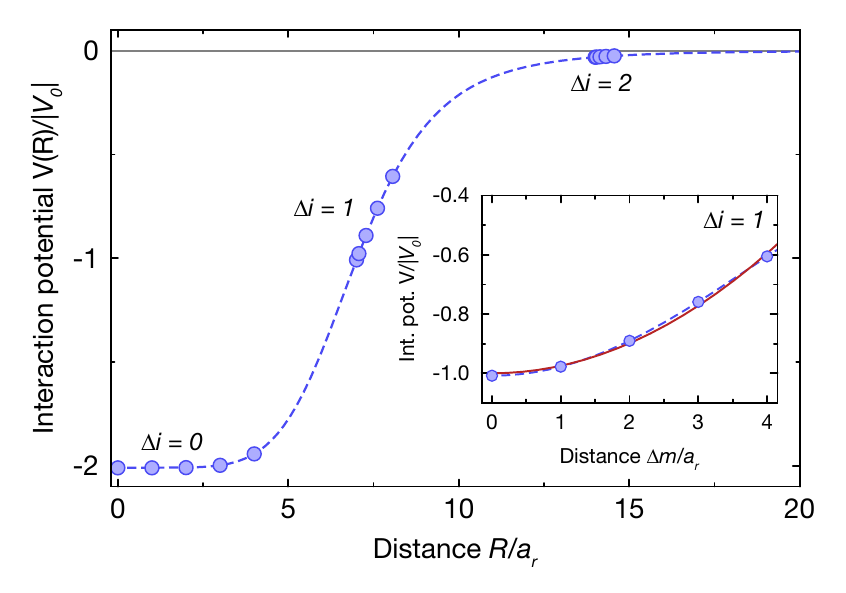}
\caption{\label{fig:dressingpotentialladder} Quadratic interactions on an asymmetric ladder for $s=2$. The isotropic Rydberg-dressed potential (dashed blue line) is sampled at different distances occurring in the ladder (blue points). Interactions between atoms in different rungs separated by $\Delta i=|i-i^\prime|$ occur in groups. The inset shows the approximate quadratic dependence for $\Delta i=1$ versus distance $\Delta m=|m-m^\prime|$ within a rung compared to a true quadratic interaction (red solid line). The parameters used are $R_c=a_l=7\,a_r$.
}
\end{figure}
While the tailored on-site potentials $\epsilon_{m,i}$ can be generated using optical potentials controlled at the single-site level~\cite{Fukuhara2013}, realizing the quadratic distance dependence of the interaction between two particles is challenging in cold atomic gases. However, they can still be realized approximately using off-resonant optical coupling of the atoms to Rydberg states. The resulting isotropic Rydberg-dressed interactions~\cite{Henkel2010,Pupillo2010} in cold atoms have recently begun to be explored in a many-body setting~\cite{Zeiher2016} and exhibit a characteristic distance dependence $V(R)=U_0/(1+(R/R_c)^6)$ for two atoms separated by a distance $R$. The saturation value $U_0$ can be tuned to be positive or negative, and the interaction range $R_c$ is set by the interactions of the coupled Rydberg states and typically reaches up to several sites in an optical lattice~\cite{SI}.

The key idea in implementing quadratic interactions in the ladder model consists in utilizing an asymmetric ladder with different lattice constants along legs and rungs respectively. In the limit of large $a_l/a_r$, the interaction potential along the rung approximately acquires the desired quadratic distance dependence for neighboring rungs with $|V_0|=|U_0|/(1+(a_l/R_c)^6)$ and $Y=6|U_0|(a_l/R_c)^6 (a_r/a_l)^2/(1+(a_l/R_c)^6)^2$. At the same time, interactions between next-nearest-neighbor rungs can be minimized, see Fig.~\ref{fig:dressingpotentialladder}, making them irrelevant for the predicted collapse shown in Fig.~\ref{fig:collapse1}. This and other imperfections as well as concrete experimental numbers are further discussed in~\cite{SI}.

A strength of the presented ladder implementation is the simple realization of models with different spin.
A natural first step would be to check the experimental feasibility of the proposal with just two legs, i.e. $s=1/2$ in Eq.~\eqref{eq:ladderH}.

The emerging spin model corresponds to the well studied spin-$1/2$ quantum Ising chain in a transverse field with the Hamiltonian

\begin{equation}
 \label{eq:isingham}
 \hat{H}=-\lambda \sum_{i=1}^{N_s}\hat{\sigma}_i^z\hat{\sigma}_{i+1}^z-h_x\sum_{i=1}^{N_s} \hat{\sigma}_i^x-h\sum_{i=1}^{N_s} \hat{\sigma}_i^z .
\end{equation}
The transverse field is realized by the tunneling of the atoms and has a strength $h_x=J/2$. Tuning $\epsilon_{\pm1/2} = \pm h - (N_s-1)(|V_0|-\lambda)/N_s$, $V_{1/2,1/2} = -|V_0|$, $V_{-1/2,1/2} = -|V_0| + 2\lambda$ is required to realize the other two terms.

 \begin{figure}[t]
 \vskip-5pt
\includegraphics[width=8.6cm]{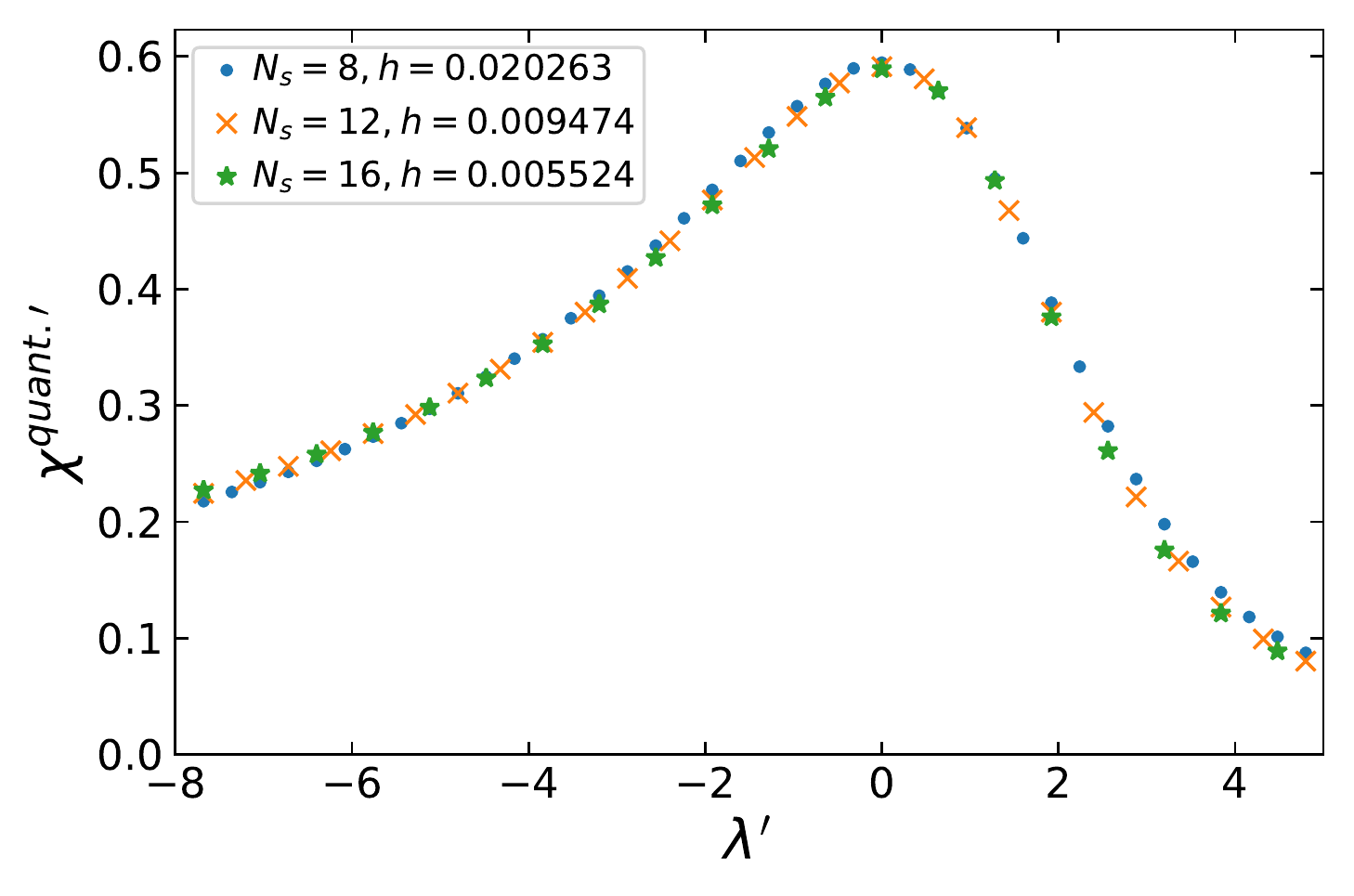}
\caption{\label{fig:coll}Data collapse for the quantum magnetic susceptibility of the quantum Ising chain with the known rescaling ${\chi^{quant.}}'= \chi^{quant.} L^{-(1-\eta)}$ versus $\lambda'=L^{1/\nu}(\lambda -1)$.
The reduced magnetic field $h^\prime = h L^{\frac{15}{8}}=1$ for all three system sizes.}
\end{figure}

Expressing all energies in units of the transverse field ($h_x=1$), this model has a second order phase transition at $\lambda = 1$ with known exponents \cite{pfeuty1970one}. As quantum simulations are still made on relatively small lattices, it is convenient to study the finite size scaling dictated by the Renormalization Group (RG) analysis of the second-order phase transition. The zero temperature magnetic susceptibility reads
 \beq
\chi^{quant.}=\frac{1}{L}\sum_{< i,j>} \langle(\hat{\sigma}_i^z -\langle\hat{\sigma}^z _i\rangle)(\hat{\sigma}^z_j-\langle\hat{\sigma}^z_j\rangle)\rangle
\end{equation}
where $\langle ... \rangle$ are short notations for $\bra{\Omega} ...\ket{\Omega}$ with $\ket{\Omega}$ the lowest energy state  of $\hat{H}$. 
The data collapse obtained with the standard RG rescalings is illustrated in Fig. \ref{fig:coll}.

The quantum Ising model has for example been quantum simulated in systems of ultracold ions~\cite{Blatt2012} and with atoms in tilted optical lattices~\cite{Simon2011}. 
New generations of D-wave machines have more versatile time-dependent capabilities. It seems possible to maintain a transverse field \cite{king18} but there are temperature effects that need to be better understood. 
Multi-mode cavity photon-mediated interactions \cite{Vaidya:2018fp} can also be used to simulate the quantum Ising model.
The possibility of extending these setups or related ones  to reproduce a multi-leg ladder 
is being investigated.

In conclusion, we have presented an experimental platform for the quantum simulation of the Abelian Higgs model in 1+1 dimension and outlined a strategy for an initial benchmark of the quantum simulator. An interesting perspective is the experimental simulation of out-of-equilibrium dynamics following a quantum quench, which promises insight into dynamics described by the LGT when inaccessible with classical computing.

{\it Acknowledgments.}
We thank I. Bloch, S. Catterall and D. Lidar for valuable conversations. 
This work was supported in part by the U.S. Department of Energy (DOE) under Award Number DE-SC0010113 (YM) and DE-SC0009998 (JUY) and by the NSF under Grant No. DMR-1411345 (SWT). JZ was supported by MPG and the European Commission (UQUAM).

\bibliography{centralmacbib2.bib}

\section*{Supplemental Material}
\section{Quadratic interactions in a ladder}
For the quantum simulation of the Abelian Higgs model in spin-$s$ truncation, we aim to implement a ladder system with quadratic interactions between particles in neighboring rungs of the ladder. The starting point is an isotropic Rydberg-dressed potential $V(R)=U_0/(1+(R/R_c)^6)$ with a cutoff distance $R_c=(C_6/2\Delta)^{1/6}$ and a saturation value $U_0=\Omega^4/8\Delta^3$ given by the laser coupling $\Omega$, detuning $\Delta$ and van-der Waals coefficient $C_6$ for the interaction between the coupled Rydberg states~\cite{Henkel2010,Pupillo2010}. Quadratic interactions can be realized in a ladder with different lattice constant $a_l$ along the legs and $a_r$ along the rungs.
To see this, we express the distance between two particles in terms of the rung and leg lattice constants and indices as 
\begin{align}
R=\sqrt{(\Delta m\,a_r)^2+(\Delta i\,a_l)^2},
\end{align}
where we have abbreviated the separations along the legs and rungs with  $\Delta i\equiv|i-i^\prime|$ and  $\Delta m\equiv|m-m^\prime|$. 
Inserting this in the interaction potential $V(R)$, we obtain for the nearest-neighbor rung with $\Delta i=1$
\begin{align}
V(R)&=\frac{U_0}{1+(\sqrt{(\Delta m a_r)^2+(a_l)^2}/R_c)^6}\\
&\approx\frac{U_0}{1+(a_l/R_c)^6}-\frac{1}{2}\frac{6 U_0(a_l/R_c)^6 (a_r/a_l)^2}{(1+(a_l/R_c)^6)^2}~\Delta m^2 \nonumber\\
&+ O\left((a_r/a_l)^4 \Delta m^4\right). \nonumber
\end{align}
For the expansion, we have assumed $(\Delta m\,a_r)/a_l\ll 1$ 
This allows us to express the potential
\begin{align}
V=-|V_0|+\frac{Y}{2} \Delta m^2
\end{align}
given in the main text in terms of the experimentally relevant parameters $U_0$, $a_l$, $a_r$ and $R_c$ by identifying
\begin{align}
Y &= \frac{6|U_0|(a_l/R_c)^6 (a_r/a_l)^2}{(1+(a_l/R_c)^6)^2}\label{eq:HiggsY}\\
|V_0| & = \frac{|U_0|}{1+(a_l/R_c)^6}.\label{eq:HiggsV}
\end{align}
For simplicity, we have hereby assumed attractive interactions and written $U_0=-|U_0|$.

As the interactions between next-nearest-neighbor rungs are assumed to be vanishing in the desired spin model in Eq.~\eqref{eq:ham}, it is important to check that they are small enough in this approximation. The next-nearest-neighbor-rung interactions (NNNRI) are obtained by just setting $a_l\rightarrow 2a_l$ in the above parameters, yielding
\begin{align}
Y^{(2)} &= \frac{6 |U_0|(2a_l/R_c)^6 (a_r/2a_l)^2}{(1+(2a_l/R_c)^6)^2}\\
|V_0^{(2)}|& = \frac{|U_0|}{1+(2a_l/R_c)^6} \\
H_{NNNRI} &=\nonumber \\ \sum_{i = 1}^{N_s - 2}\sum_{m,m^\prime=-s}^{s} & \left(-|V_0^{(2)}| + \frac{Y^{(2)}}{2} \Delta m^2\right) \hat{n}_{m,i} \hat{n}_{m^\prime,i+2}.
\end{align} 
In order to continue the discussion, it is now helpful to gain some intuition for where the quadratic dependence is a good approximation to the Rydberg-dressed potential. One exemplary Rydberg-dressed potential is depicted in the main text in Fig.~\ref{fig:dressingpotentialladder}. As one can see in this example, the points for the different $\Delta i$ are grouped. This is generally necessary to achieve a large ratio between $\Delta i=1$ and $\Delta i=2$, i.e. a suppression of NNNRI. Furthermore, the $\Delta i=1$ points should be located close to the most linear part of the potential $R\approx R_c$. This can be understood from the geometric argument that in the limit $a_r/a_l\rightarrow0$, where the quadratic approximation of $R$ in $\Delta m$ works best, we sample a linear potential with approximately quadratically spaced points. 
At the same time, however, pushing $a_r/a_l$ towards zero squeezes the points closer together, effectively reducing the interaction $Y$. This is also directly obvious from the quadratic dependence of $Y$ on $a_r/a_l$ shown in Eq.~\eqref{eq:HiggsY}. To summarize, one has to compromise between maximizing $Y$, minimizing $Y^{(2)}$ and optimal quadratic dependence within a rung.

Qualitatively, Fig.~\ref{fig:dressingpotentialladder} and our previous discussion indicate to work close to the regime $R_c\approx a_l$. Setting for simplicity $R_c=a_l$, we can estimate the ratio between nearest-neighbor and next-nearest-neighbor rung interactions as (approximating $2^6+1\approx 2^6$)
\begin{align}
|V_0|      &= \frac{1}{2}|U_0|\\
|V_0^{(2)}|&=\frac{1}{64}|U_0|\\
Y       &=\frac{3}{2}|U_0|\left(a_r/a_l\right)^2\label{eq:Y_Rc_eq_al}\\
Y^{(2)}  &=\frac{3}{128}|U_0|\left(a_r/a_l\right)^2.
\end{align}
This shows that NNNRI are suppressed by approximately $|Y|/|Y^{(2)}|=2^6=64$ in this regime and hence only reach up to $1.5 \%$ of the nearest-neighbor-rung interactions (NNRI). Furthermore, it should be noted that already small changes in $a_l/R_c$ influence the exact numbers due to the strong scaling with $(a_l/R_c)^6$, which allows for considerable flexibility in fine-tuning the implementation.
For example, increasing $a_l/R_c$ slightly above unity suppresses the NNNRI to values below $1\%$, such that they can easily be tuned to be irrelevant for the collapse displayed in Fig.~\ref{fig:collapse1}.

\begin{figure}[t]
\includegraphics[width=8.6cm]{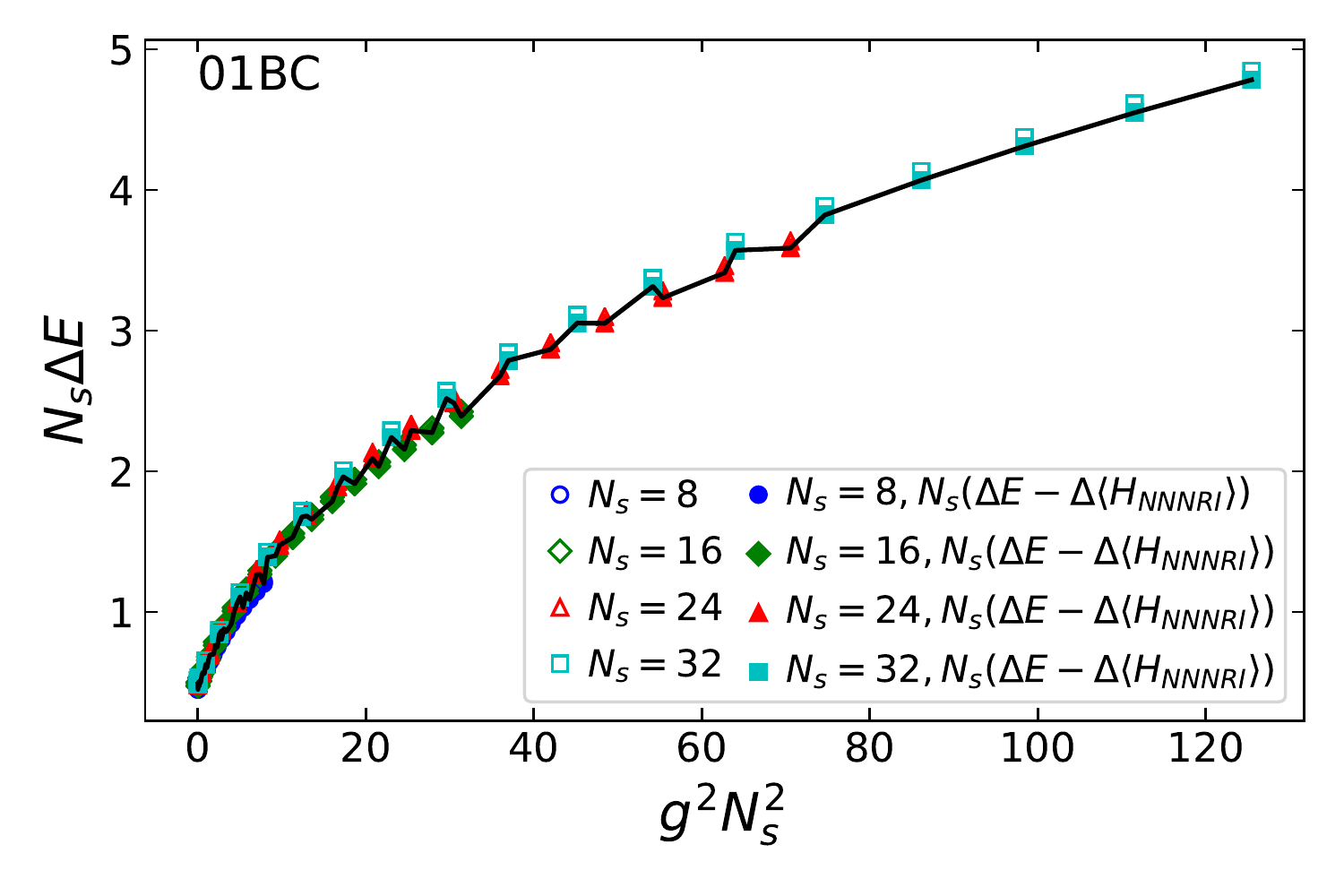}
\includegraphics[width=8.6cm]{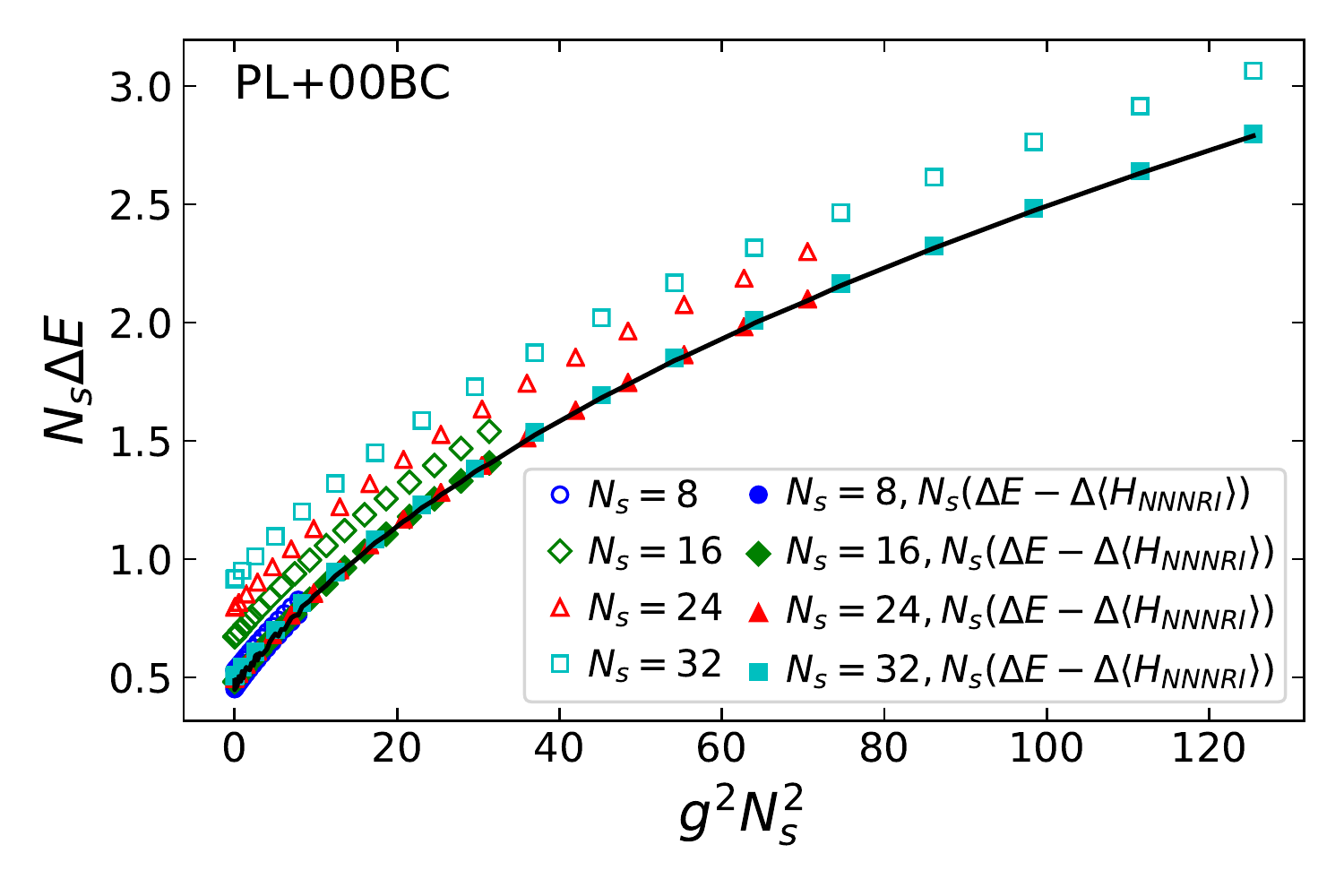}
\caption{\label{fig:collapseladderwithnnri} $N_s \Delta E$ obtained from the ladder Hamiltonians with NNNRI $1.5\%$ of NNRI (open markers). The ladder systems which are mapped to the spin Hamiltonians with and without 01BC (top) are much more robust to NNNRI than those which are mapped to the spin Hamiltonians with and without a Polyakov loop (bottom). We can correct the ground state energy by subtracting the ground state expectation value of NNNRI from it (filled markers). The solid lines are the collapse lines from the spin Hamiltonians.
}
\end{figure}

Fig.~\ref{fig:collapseladderwithnnri} shows $N_s \Delta E$ obtained from the Hamiltonian of the ladder system with NNNRI $1.5 \%$ of NNRI by DMRG. The data collapse is robust to NNNRI for the ladder systems which are mapped to the spin Hamiltonians with and without 01BC, while it is broken dramatically by such a small NNNRI in the ladder systems which are mapped to the spin Hamiltonians with and without a Polyakov loop. In the $O(2)$ limit, NNNRI doesn't change the critical property for the former, while it would open a gap to the gapless phase for the latter, which has been confirmed numerically (not shown here). However, based on 1st order perturbation theory, we can correct the ground state energy by subtracting the expectation value of NNNRI, $\langle H_{NNNRI} \rangle$, from it. The corrected data sets collapse perfectly onto the same lines of collapse in Fig.~\ref{fig:collapse1}. Experimentally, we can measure the density-density correlations between next-nearest-neighbor rungs $\langle \hat{n}_{m, i} \hat{n}_{m^\prime, i+2} \rangle$ and extract $\langle H_{NNNRI} \rangle$ to do the same correction.

\subsection{Imperfections due to decoherence}
Experimentally, the achievable Rydberg-dressed interaction strength $U_0$ is limited via the effective decay rate $\gamma_{\mathrm{eff}}$ of the dressed state due to the coupling to a Rydberg state with a limited lifetime~\cite{Henkel2010,Zeiher2016}. This leads to decoherence in the time evolution of the system not captured by purely Hamiltonian dynamics described by Eq.~\eqref{eq:ladderH}. Generally, both the Rydberg-dressed interaction strength as well as the effective decay rate increase with larger Rydberg-state admixture. One can quantify the coherence in terms of a ``figure of merit" $Q=|U_0|/\gamma_{\mathrm{eff}}$, which expresses the average number of coherent cycles per decay event.\\
For simulating the Abelian Higgs model in the spin representation, the additional constraint of realizing quadratic interactions effectively reduces the usable interactions. Considering Fig.~\ref{fig:dressingpotentialladder}, this is directly obvious as the strongest interactions ($\Delta\,i=0$) are not sampled for a single atom per rung. A useful figure of merit is given by comparing the quadratic part of the interaction Eq.~\eqref{eq:HiggsY} with the effective decoherence,
\begin{align}
Y/\gamma_{\mathrm{eff}} &= 6\frac{(a_l/R_c)^6 }{(1+(a_l/R_c)^6)^2}(a_r/a_l)^2|U_0|/\gamma_{\mathrm{eff}}\nonumber\\
&=6\frac{(a_l/R_c)^6 }{(1+(a_l/R_c)^6)^2}\left(a_r/a_l\right)^2 Q.
\end{align}   
This shows that realizing an optimal quadratic dependence, which requires ideally $a_r/a_l\rightarrow 0$, as well as maximally suppressing the NNNRI, which requires $R_c/a_l\rightarrow0$,  conflicts  with the interaction-to-decoherence ratio. Below, we outline a parameter regime showing that the experimental implementation still seems feasible.

\section{Measurement of the energy gap}
In order to experimentally measure the energy gap displaying the universal scaling (see Fig.~\ref{fig:collapse1}), it is necessary to first prepare the ground states of the respective models. Their mean energy $\langle E\rangle$, with $\langle\ldots\rangle$ denoting the quantum average, can then be reconstructed by two sets of measurements. In a first set of measurements, the atomic distributions are immediately frozen  by switching off interactions and tunneling at the same time. The contributions of on-site potentials, $\sum_{i=1}^{N_s}\sum_{m=-s}^{s}\epsilon_{m,i}\langle\hat{n}_{m,i}\rangle$, and interactions, $\sum_{i,i^\prime=1}^{N_s}\sum_{m,m^\prime=-s}^{s}V_{m,m^\prime,i,i^\prime}\,\langle\hat{n}_{m,i} \hat{n}_{m^\prime,i^\prime}\rangle$, to the mean energy can then be extracted from the measured atomic distribution by measuring the mean local density $\langle\hat{n}_{m,i}\rangle$ and density-density correlations $\langle\hat{n}_{m,i} \hat{n}_{m^\prime,i^\prime}\rangle$ respectively. Both are accessible in a quantum gas microscope with local detection~\cite{Endres2013a}. In a second set of measurements, the contribution of the tunneling term to the mean energy has to be determined. This amounts to extracting first-order correlations of the form $\langle\hat{a}_{m,i}^\dagger \hat{a}_{m+1,i}\rangle$, which has recently been demonstrated for atoms in optical lattices employing Talbot interferometry~\cite{Santra2017}. The reconstruction of the energy relies on knowledge of the on-site potentials, interactions and tunnel coupling along the rungs, each of which can be calibrated in independent measurements.  
\section{Experimental numbers}
In the following, we give some concrete experimental figures to underline the feasibility of realizing the desired spin models. Hereby, we assume tunneling along the rungs with strength $J/h=100\,$Hz, which is e.g. readily achieved for rubidium in optical lattices~\cite{Weitenberg2011a}.
\subsection{Spin-$s$ model with quadratic interactions}
First, we focus on the feasibility of reaching the parameter regime for the ladder implementation of the spin-$s$ truncation of the Abelian Higgs model to study the collapse displayed in Fig.~\ref{fig:collapse1}.
Aiming at $J=Y$ ($X=1$ in Fig.~\ref{fig:collapse1}), the required interaction strength for $R_c=a_l\approx7\,a_r$ is $|U_0|/\hbar\approx 2\pi\times 3.3\,$kHz (using Eq.~\eqref{eq:Y_Rc_eq_al}).
Assuming a coupling strength of $\Omega/\hbar=2\pi\times100\,$MHz to the Rydberg state, which is within reach in future, specialized experiments, the desired interaction $|U_0|$ and hence $Y$ can be achieved with a detuning of $\Delta/\hbar\approx2\pi\times 1560\,$MHz. 
The interaction strength between two Rydberg atoms with $80P_{3/2}$, measured by the van der Waals dispersion coefficient, is $C_6\approx5500\,\mathrm{GHz}\,\mu\mathrm{m}^6$. As a result, for the chosen detuning, the cutoff distance is $R_c=(C_6/2\Delta)^{1/6}\approx3.5\,\mu$m and hence an experimentally realistic lattice spacing of $a_r\approx 500\,$nm would lead to $R_c/a_r\approx7$.
For the lifetime $\tau=250\,\mu$s of the admixed Rydberg state $80P_{3/2}$, the dressed state acquires an effective decay rate of $\gamma_{\mathrm{eff}}\approx 4\,\mathrm{s}^{-1}$, resulting in $J/\gamma_{\mathrm{eff}}\approx25$.
Tuning to larger values of $J/Y$ by reducing $Y$ to study the collapse shown in Fig.~\ref{fig:collapse1} improves this ratio by a factor $\sqrt{J/Y}$ for constant $\Delta$ due to smaller Rydberg-state admixture.\\
A direct increase in the figure of merit should be possible by using lighter elements like potassium or lithium, for which larger tunneling rates are feasible. This allows for working at a larger Rydberg-state admixture, which is favorable to increase the quality factor. Furthermore, improvements are possible by increasing Rabi couplings to Rydberg states, by implementing more advanced dressing schemes~\cite{Bijnen2015} or in future experiments in a cryogenic environment, where the Rydberg lifetime is expected to increase (to $1.2\,$ms for $80P$) due to suppression of black-body-radiation-induced decay. Combining these steps, we estimate a possible increase in the figure of merit by well above one order of magnitude.\\

\subsection{Spin-$1/2$ Ising model}
The spin-$1/2$ Ising model (Eq.~\eqref{eq:isingham}) discussed in the main text constitutes a first, easier step for implementing the described ladder systems. Contrary to the more complex spin-models, in this case no asymmetric ladder is required, i.e. $a_r=a_l$. Keeping the condition $R_c=a_l$, in this case $\lambda\approx U_0/4$ using the same definition as in Eq.~\eqref{eq:isingham}. 
For a Rabi-coupling of $\Omega/\hbar=2\pi\times100\,$MHz and a detuning of $\Delta/\hbar\approx2\pi\times 1560\,$MHz, the accessible spin interaction strength is $\lambda/\hbar=2\pi\times825\,$Hz. A tunneling strength of $J/\hbar=2\pi\times100\,$Hz along the rungs translates to a transverse field of $h_x/\hbar=2\pi\times50\,$Hz, such that $\lambda/h_x\approx16.5$. For a coupling to the Rydberg state $80P_{3/2}$ with an effective decay rate of $\gamma_{\mathrm{eff}}\approx 4\,\mathrm{s}^{-1}$ for the above admixture parameters, the figure of merit in this regime becomes $h_x/\gamma_{\mathrm{eff}}\approx12.5$. Towards stronger transverse fields, the figure of merit increases proportional to $\sqrt{h_x/\lambda}$, such that it reaches up to $50$ in the critical region $\lambda/h_x=1$. Furthermore, it can be improved by the same means quoted above for the more complex spin model.
\end{document}